# Transport phenomena of TiCoSb: Defects induced modification in structure and density of states

S. Mahakal[1], Diptasikha Das[2], Pintu Singha[3], Aritra Banerjee[3], S. Chatterjee[4], Santanu K. Maiti[5], S. Assa Aravindh[6] and K. Malik*[1]

[1] *Department of Physics, Vidyasagar Metropolitan College, Kolkata-700006, India.*
[2] *Department of Physics, ADAMAS University, Kolkata-700126, India.*
[3] *Department of Physics, University of Calcutta; Kolkata-700009, India.*
[4] *UGC-DAE Consortium for Scientific Research, Kolkata Centre, Kolkata 700 098, India.*
[5] *Physics and Applied Mathematics Unit, Indian Statistical Institute, 203 Barrackpore Trunk Road, Kolkata-700 108, India.*
[6] *Nano and Molecular Systems Research Unit (NANOMO), University of Oulu, FIN-90014, Finland.*

TiCoSb1+x (x=0.0, 0.01, 0.02, 0.03, 0.04, 0.06) samples have been synthesized, employing solid state reaction method followed by arc menting. Theoretical calculations, using Density Functional Theory (DFT) have been performed to estimate band structure and density of states (DOS). Further, energitic calculations, using first principle have been carried out to reveal the formation energy for vacancy, interstitial, anti-site defects. Detail structural calculation, employing Rietveld refinement reveals the presence of embedded phases, vacancy and interstitial atom, which is also supported by the theoretical calculations. Lattice strain, crystalline size and dislocation density have been estimated by Williamson-Hall and modified Williamson-Hall methods. Thermal variation of resistivity [ρ(T)] and thermopower [S(T)] have been explained using Mott equation and density of states (DOS) modification near the Fermi surface due to Co vancancy and embedded phases. Figure of merit (ZT) has been calculated and 4 to 5 times higher ZT for TiCoSb than earlier reported value is obtained at room temperature.

## I. Introduction

Half-Heusler (HH) alloys are drawing attention of the researcher for fascinating and unconventional physical properties along with non-toxic, good thermal stability and high mechanical strength etc.[1,2,3,4,5] Worldwide resurgence is going on to harnessing interesting multinary properties with unusual electronic and magnetic behaviour of HH alloys into device applications.[4,6,7] Vacancies, present in Heusler compound and valence electron count (VEC) are sole responsible for novel electronic and magnetic properties of this group of materials. Crystal structure of the HH alloys depends on the constituent elements and changes with composition viz, ZrNiSb, HfNiSb and ScCoSb.[8,9] MCoSb [M=Ti, Zr, Hf], HH alloys are inter-metallic and crystallize as MgAgAs cubic structure with space group $F\bar{4}3m$.[10,11,12,13] It may be considered as three inter penetrating face-centered cubic (fcc) lattice of almost identical unit cell volume. MCoSb are in the category of XYZ type HH alloy, where X is more electropositive than transitional metal Y and main group Z element.[13] For MCoSb, one sublattice consists of lower valent transition metel, M along with sp atoms Sb; and another one has been formed by higher valence transition metal, Co and vacancies. The materials exhibit semiconducting character in spite of constitute element being metal.[12,13] The structure may be considered as FCC lattice with X at Octrahedral site (4a= (0, 0, 0)). The structure is combination of rock salt (XZ) and Zinc blande (YZ) sub-lattice with Y is at 4c (¼, ¼, ¼) and Z at 4b (½, ½, ½).[13] The 4d (¾, ¾, ¾) site in $F\bar{4}3m$ is filled for the XY$_2$Z Heusler alloy.[13] The tetrahedral site 4d may be referred as the interstitial site for the XYZ structure. A typical schematic diagram of TiCoSb unit cell has been presented in Fig. 1. Wyckoff position of Ti, Co and Sb may also be considered at 4c (1/4,1/4,1/4), 4a (0,0,0), 4d (3/4,3/4,3/4) and vacant 4b (1/2,1/2,1/2) for the MgAgAs type structure.[14] TiCoSb unit cell structure with vacant 4d is more stable i.e., first case.[14] However, the constituent atoms and structure strongly corroborated with the electronic and transport properties of HH alloy. MCoSb is 18 VEC type materials and Fermi level is just above valence band.[15,16,17,18] Semiconducting properties in 18 VEC HH alloy arise due to the transfer of electron from the most electro positive X to the less electro positive Y and Z elements.[13] However, HH alloys having VEC=18 may be understood from the Zintal chemistry framework. Close valence cell configuration is satisfied due to the transfer of valence electron from electro positive $X^{n+}$ to covalent $YZ^{n-}$ sub-lattice.[13] It is crucial to point out that electro negativity of X, Y and Z site atoms lies in the range 1.2-1.7, 1.8-2.4, 1.7-2.2 respectively.[19] And close cell configuration for the Y and Z atoms i.e. $d^{10}$ and $s^2p^6$ is achieved for the HH alloys.

Research in thermoelectric (TE) material is emerging as fascinating topic due to potential alternative source of energy and peculiar physics involved within the technology.[4, 5] Thermoelectricity converts thermal energy into electrical energy and vice versa. Till now, several systems have been studied as highly efficient TE materials viz, PbTe based materials (ZT~1.5), chalcogenides [$Bi_2Te_3$ (ZT~1), $Sb_2Te_3$ (ZT~1)], $Mg_2Si$ (ZT~1.4) and oxide TE materials (ZT~2.7) etc.[13] However, applications are limited by the poor chemical, mechanical strength and thermal stability.[13] The choice of material, TiCoSb is potential to be used as TE material at mid-temperature.[20] MCoSb [M= Ti, Co, Sb] is a potential p-type TE material and resurgence are going on to enhance the efficiency by reducing thermal conductivity ($\kappa$).[10] Large Seebeck coefficient (S(T)) and electrical conductivity ($\sigma$) due to d-d bonding near the Fermi surface, made them attractive and potential as TE material. TiCoSb is para magnetic and temperature independent.[21] Terada et al. have reported that it shows ferromagnetic behaviour due to different ferroelectric phase precipitation in HH alloy matrix.[21] However, Iso-electronic alloying, defects and disorder in the structure influence the electronic, transport properties and concomitantly the TE properties of the HH alloy.[22] TiCoSb shows lower band gap ~0.95eV amid the other iso-electronic alloys, ZrCoSb and HfCoSb.[23] Introduction of point defect at X site and reduction of $\kappa$ enhances the ZT~1 for n-type and p-type (MNiSn and MCoSb) HH Alloys.[24] Nano inclusion and complex iso-electronic alloying are also important routes to enhance the efficiency of a TE material.[10, 25] Further, $\kappa$ may be significantly reduced by point defect and disorder through phonon scattering. Iso-electronic alloying not only modifies electronic structure but also reduces $\kappa$ by introducing point defects in the HH structure.[25] However, nominal amount of impurities with 19 VEC HH alloy strongly affect the TE properties of the HH alloy.[26] But, scattering of carrier with defect and disorder concomitantly reduces $\sigma$.[27] Xia et al. have described that short range order along with long rage periodicity significantly enhance carrier conductivity accompanying the low $\kappa$.[28] Further, disorder also strongly affects the band structure of n-type MNiSn based HH alloy.[29, 30, 31] Effect of iso-electronic alloying and nano-inclusion in TE properties have been studied widely.[10, 25, 32] However, investigation on consequences of defect and disorder on crystal, electronic structure and transport properties of TiCoSb is limited.[33]

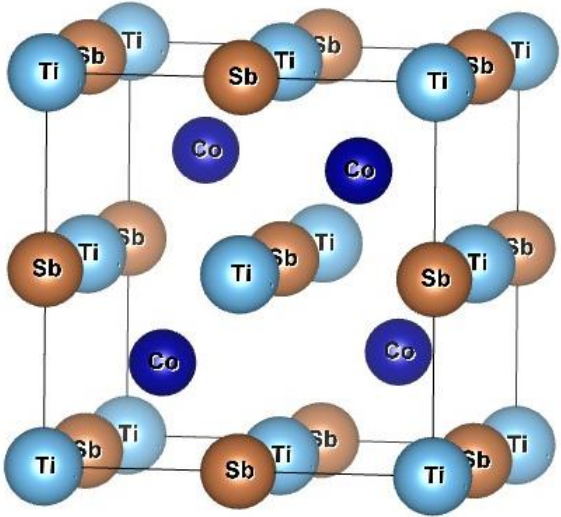

**Fig. 1**. (Color online). Schematic view of the arrangements of Ti, Co, Sb atoms in TiCoSb half-Heusler alloy. Sky blue spheres represent Ti atoms, brown spheres represent Sb atoms and navy blue spheres represent Co atoms.

Phase segregation in the HH alloy matrix is one of the important aspects of the TE study. In order to avoid the presence of second phases and defects, several methods have been implemented to synthesize TiCoSb HH alloy.[34, 35, 36] Sometimes, defects are favourable to enhance TE properties through modification in electrical and thermal transport properties.[34, 35] However, ZT has been reduced due to segregation of secondary phase viz, CoSb in TiCoSb matrix.[37] M. Assad et al. have reported that embedded phases in doped TiCoSb enhance the ZT due to a reduction in $\kappa_L$.[34] C. S. Birkel et al. have synthesized TiCoSb in two different ways, mechanical alloying and microwave synthesis. And it has been observed that the second phase causes a reduction in TE properties.[35] Arc melting is one of the common techniques to synthesis HH or FH alloys. But the difference in melting points of constituent elements in the stoichiometry causes a source of defect/segregation in the HH alloy.[34, 36] In order to avoid loss due to evaporation, there are instantses to use extra Sb in TiCoSb stoichometry.[37] However, role of defects and second phases on the TE properties of TiCoSb require a systematic study. There is no meticulous investigation for inclusion of extra Sb during synthesis of TiCoSb and effect on TE properties due to the complex interplay of vacancy and embedded phases. Furthermore, this study may enlighten the source of defects, embedded phases in the synthesized matrix and correlation with the transport properties of TiCoSb HH alloy.

Here we have reported, the role of defects and precipitated phase on electronic structure and transport properties of polycrystalline TiCoSb alloy, synthesized by solid state reaction followed by arc melting. In-depth structural characterizations have been performed using Rietveld refinement method from the x-ray diffraction (XRD) data. Occupancy and precipitated phase have been revealed from XRD. Vacancy in the synthesized materials has been shown schematically. Further, electronic band structure, DOS and formation energy of TiCoSb have been estimated using density functional theory (DFT) calculation, considering super cell structure. Effects of vacancy, interstitial defect and anti-site defects have been also included in DFT calculations. Temperature dependent resistivity ($\rho(T)$) and S(T) measurements have been carried out at low temperature down to 10K. Electron-electron (e-e) and electron-phonon (e-ph) coefficients have been estimated from the thermal variation of $\rho(T)$ and S(T) data. First time in-depth structural study using the refinement and correlation with low temperature transport properties have been presented in this article. Further, attempts have been taken for theoretical calculation of band structure, density of states (DOS) and total energy per atom. Bipolar conduction has been found, according to the results obtained from transport properties and correlated with the structural parameters, evaluated using Rietveld refinement. DOS has been tuned by modification in structural parameter and correlation with $\rho(T)$, S(T) data have been illucidated, using mott equation. Theoretical and experimental data along with the estimated parameters have been corroborated in this endeavour.

In our present work, we have considered TiCoSb as a functional element for suitable energy conversion. TiCoSb1+x (x=0.0, 0.01, 0.02, 0.03, 0.04, 0.06) has been synthesized to study the role of defects and embedded phases in stuctural, transport and electronic properties. The key findings of the present study are (i) Co vacancies in TiCoSb unitcell increase with Sb concentration, (ii) minute amount of CoTi and CoSb embedded phases have been revealed and supported by theoretical study, (ii) embedded phases, dislocation density and strain are minimum, correspondingly maximum TiCoSb phase is observed in TiCoSb1.02, (iii) modification in thermal variation of resistivity and thermopower is related with DOS due to Co vacancies, (iv) inclusion of 2% Sb in TiCoSb stoichimetry during synthesis is sufficient to compensate the weight loss, and (v) 4 to 5 times higher ZT than previously divulged for TiCoSb at room temperature has been obtained.

This paper is organized as follows. Section II describes precise steps for sample synthesis and characterizations to analyse the experimental results. Computational details for theoretical calculations have been presented in Sec. III. All the theoretical and experimental results have been critically scruitnized and discussed in Sec. IV. Finally, conclusion and correlation have been drawn on the basis of theoretical and experimental results in Sec. V.

## II. Experimental details

Polycrystalline TiCoSb1+x (x=0, 0.01, 0.02, 0.03, 0.04 and 0.06) were synthesized by solid state reaction. Weighted Ti (purity 99.998%; Alfa Aesar, UK,) Co and Sb (each of purity 99.99%; Alfa Aesar, UK ) were arc melted at vacuum and ingots were melted several times to obtain homogeneous alloy. The arc melted ingots were sealed in quartz ampoule under pressure $10^{-3}$ m.bar for further treatment. Evacuated quartz ampoules were annealed at 1173 K for five days and then cooled down to room temperature at cooling rate $10^0$C/hour. The prepared ingots were cut in rectangular shape of dimension (1.5mm x 2mm x 5mm) measurement of transport properties. In order to carry out XRD measurements, prepared ingots were pulverised into fine powders. Structural characterization of the synthesized host phases i.e. TiCoSb, TiCoSb1.01, TiCoSb1.02, TiCoSb1.03, TiCoSb1.04, TiCoSb1.06 have been carried out by XRD using Diffractometer model: X'Pert Powder, PANalytical using Cu-$K_\alpha$ radiation of wavelength 0.15406 nm. XRD has been carried out at room temperature in the range $20^0 \leq \theta \leq 80^0$ in $\theta$-2$\theta$ geometry. The Williamson-Hall and Modified Williamson-Hall methods have been employed to estimate lattice strain ($\varepsilon$), crystal size (d) and dislocation densities ($N_D$) of the synthesized samples using the XRD data.[38, 39] Further, in-depth structural analysis has been carried out by Rietveld refinement technique using FullProf software.[40] Refinement analysis of the Bragg scattered TiCoSb1+x phases have provided exhaustive information on the long-range order in the host phase along with defect and disorder in the synthesized HH alloy structure. Further, mix-phase has been introduced during Rietveld refinements to investigate the presence of defect and disorder in the TiCoSb matrix. $\rho$(T) measurements of the synthesized polycrystalline alloys have been carried out using conventional four probe method down to 10K. S(T) measurements have been performed down to 20K, employing standard differential technique.[41, 42, 43] TiCoSb is a HH alloy with a vacant 4d (3/4, 3/4, 3/4) and prone to antisite/interstial defects. The defects have been also correlated with transport properties through modifications in DOS of the synthesized samples. Embedded phases and defects in TiCoSb significantly alter the transport properties through carrier scattering and modification in electronic structure.

**TABLE I.** The simulated total energy/atom for supercell containing vacancies, anti-sites and interstials.

| | TiCoSb | Vacancy | | | Interstial | | | Antisite | |
|---|---|---|---|---|---|---|---|---|---|
| | | $V_{Ti}$ | $V_{Co}$ | $V_{Sb}$ | Ti | Co | Sb | Ti-Co | Ti-Sb |
| Energy (eV) | -6.96 | | | | | | | | |
| | | -6.72 | -6.77 | -6.76 | -6.94 | -6.93 | -6.78 | -6.82 | -6.86 |

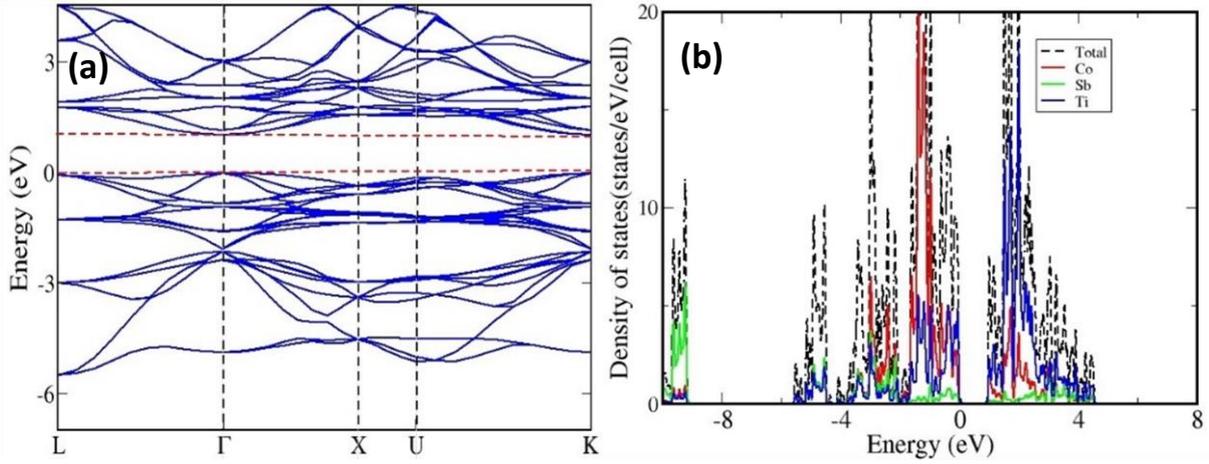

**Fig. 2.** (Color online). Electronic (a) band structure and (b) DOS of TiCoSb half-Heusler alloy.

In order to corroborate the experimental findings, DFT calculations have been carried out using the plane wave pseudo potential code Vienna Ab initio Simulation Package (VASP).[44]

## III. Computational Details

We have carried out DFT calculation for the TiCoSb using the plane wave pseudo potential code VASP.[44] To expand the plane waves included in the basis set kinetic energy cut off of 520 eV has been used. A generalized gradient approximation (GGA) has been employed in the Perdew Burke Erzenhof (PBE) formalism.[45, 46] The atomic positions have been relaxed by using energy and force tolerance of $10^{-6}$ eV and 0.001 eV/A, respectively to minimize the Hellmann-Feynmann forces. An fcc unit cell of TiCoSb has been optimized and relaxed lattice parameter of 5.88Å has been found. A Monkhorst Pack K grid of 4 x 4 x 4 has been utilized for the brillouing zone integration.

## IV. Results and Discussion

The band structure and DOS have been presented in Fig. 2(a) and Fig. 2(b) respectively. A direct band gap of about 1.03 eV at the gamma point has been obtained. The DOS calculation has shown that the valence band maximum and conduction band minimum have mainly constituted from the contribution of Co and Ti orbitals.

We have simulated a super-cell of TiCoSb, containing 48 atoms using the optimized lattice parameter of unit cell. In this super-cell, vacancies, interstitial and anti-site have been introduced. The total energy per atom of these super-cells have been presented in TABLE I. The energetic calculations have shown that the super-cell containing Co vacancy is more stable than that of Ti and Sb. Further, the system with Ti in interstitial position is more stable than Co and Sb. We have also carried out calculations by introducing Co and Sb in Ti position and found that the latter is more stable. The total energy per atom, calculated for TiCoSb super-cell with 48 atoms has also been shown for comparison. It has shown that, having an interstitial Ti atom is closer in stability to the perfect structure.

Fig. 3 represents XRD pattern of TiCoSb1+x (x=0.0, 0.01, 0.02, 0.03, 0.04, 0.06) samples, synthesized by solid state reaction method. XRD has shown that the bulk matrixes of the synthesized samples have crystallised in a cubic MgAgAs type structure. All the diffraction peaks have been indexed to the corresponding structure and space group ($F\bar{4}3m$, No. 216) [Fig. 3]. Effect of Sb concentrations on the most intense peak (220) has been demonstrated in Fig. 3 (inset) (220). XRD peaks shifts towards higher 2θ angle for 0.0≤x≤0.04, indicates reduction in unit cell volume. However, a large shift of (220) peak towards lower angle has been observed for x=0.06. Any extra impurity peak has not been found in the XRD detection level. Full Width at Half Maxima (FWHM) of (220) peaks has been estimated and presented in Fig. 4. FWHM of the synthesized samples decrease for $0.0 \leq x \leq 0.02$, indicates enhancement of crystal quality as HH alloy. Increase in Sb i.e., for $0.02 < x \leq 0.06$ increase in FWHM represents degradation of crystal quality. However, structural disorder, defects and degradation of crystal quality may arises owing to the inherent off-stoichiometry and embedded phase.

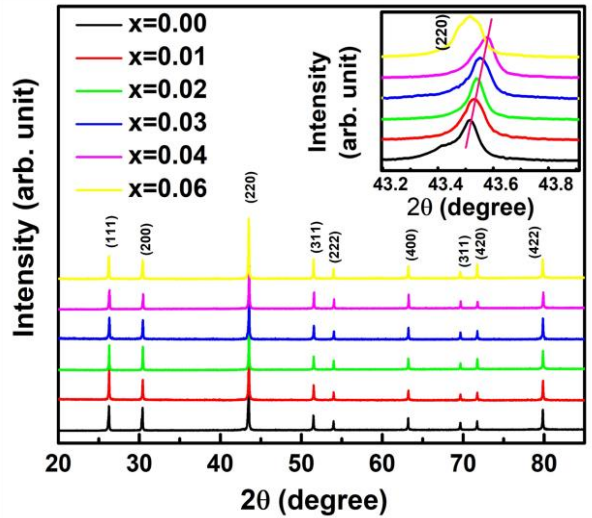

**Fig. 3.** (Color online). X-ray diffraction pattern of synthesized TiCoSb1+x (x=0.0, 0.01, 0.02, 0.03, 0.04, 0.06) polycrystalline materials (Insets shows position of highest intense peak (220) peak).

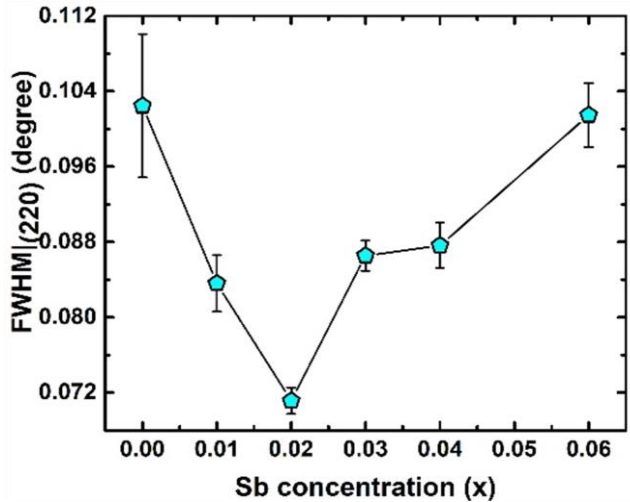

**Fig. 4.** (Color online). Variation of Full Width at Half Maxima (FWHM) with Sb concentration for TiCoSb1+x (x=0.0, 0.01, 0.02, 0.03, 0.04, 0.06). FWHM have been obtained by fitting the highest intense x-ray diffraction peak (220).

In order to estimate in-depth structural parameters, Rietveld refinement analysis has been carried out using Foolproof software.[40] Further, the presence of embedded phases has been revealed by introducing mix-phase Rietveld analysis. The CoTi and CoSb phases have been incorporated during Reitveld refinement, which improves the fitting quality, reduced $\chi^2$ (goodness of fit). However, reports on in-depth structural analysis of TiCoSb using Rietveld analysis are limited. But till date, no effort has been made to reveal the effect of Sb in structural and transport properties of TiCoSb alloy, using mix-phase Rietveld refinement. The experimental XRD patterns along with the theoretical fitted curve after Rietveld analysis have been given in Fig. S1 [supplementary information]. Lattice parameter 'a' of the synthesized samples has been estimated and a ~ 0.5883 nm has obtained for TiCoSb1+x ($0 \leq x \leq 0.06$) samples. Value of 'a' parameter is in good agreement with theoretically obtained value. Similar value of 'a' parameter has also been reported by Webster and Ziebeck (a = 0.5884 nm).[47] Unit cell volumes of the synthesized samples have been estimated and presented in Fig. 5(a). Unit cell volume gradually decreases with increasing Sb concentration for 0<x<0.04 and increases for x=0.06, corroborated with (220) peak shift in the XRD pattern (Inset Fig. 3).

In-depth mix phase Rietveld analysis has revealed the presence of a minute amount of CoTi and CoSb phases in the synthesized TiCoSb1+x matrix. Sb concentration-dependent weight percentage (wt%) of all the existing phases in the synthesized TiCoSb samples have been depicted in Fig. 5(b). wt% of CoTi phase decreases and CoSb phase increases gradually with Sb concentration in TiCoSb samples. A maximum TiCoSb phase as high as ~99% has been achieved for x = 0.02. Phase segregations have been frequently observed in ternary intermetallic, TiCoSb-based HH alloys due to large differences in specific gravity and boiling points (Sb : 1,587 °C, Ti : 1,668 °C, Co: 1,495°C) of the constituent elements.[24] Evaporation of Sb during the arc melting of weighted elements in TiCoSb samples is one of the key reasons for the segregation of phases in the HH alloy host matrix.[48] Deviation from the chemical composition of TiCoSb during synthesis may lead to the presence of CoTi-embedded phases and structural disorder in the prepared samples.[48] A meticulous investigation has revealed that the variation of wt% of TiCoSb, CoTi, CoSb with Sb concentration is associated with a slope change at x = 0.02. wt% of the host phase, TiCoSb rapidly decreases after x = 0.02. Further, systematic Co vacancy in the synthesized TiCoSb samples has been revealed using the Rietveld refinement and a schematic representation has been depicted in Fig. 6 by employing Vesta software.[49] Co vacancy gradually increases with Sb concentration in TiCoSb1+x (0<x<0.06). The decrease in lattice volume of the synthesized sample may be related to (0<x<0.04) increasing Co vacancy in the system. Das et al. have reported that Te vacancy-mediated point defects in $Sb_2Te_3$ TE material cause a decrease in unit cell volume.[50] However, a jump in lattice volume for the TiCoSb1.06 sample has been observed. Primarily, Co vacancies have created in TiCoSb system as observed from Rietveld refinement analysis and CoSb phases have formed in reaction with extra Sb. Further, the increase of Sb in TiCoSb may cause the presence of interstitial atoms and anti-site defects simultaneously. There are two possibilities arise due to the increase of Sb in TiCoSb, Sb atom either goes into the interstitial site i.e., 4d (3/4, 3/4, 3/4) or simultaneously, Ti atom goes into the interstitial vacant 4d (3/4, 3/4, 3/4) position and Sb atom occupies Ti site i.e., Sb-Ti($Sb_{Ti}$) anti-site defect. According to the theoretical calculations, Ti at the interstitial position is more favourable than other configurations, $Sb_{Ti}$ anti-site is more stable than other anti-site defects. Interstitial Ti atoms in TiCoSb unit cell causes an increase in lattice volume for x=0.06 and decrease in CoTi embedded phases in the TiCoSb matrix (Fig. 5). This result has been also corroborated with the peak shift in the XRD pattern, Fig. 3. XRD peak (220) shifts towards a higher angle for 0≤x≤0.04 compared to the TiCoSb1.06 sample [Fig. 3(inset)] owing to the collective diffraction from the embedded (segregated) phases and decreasing disorder, suggests that the lattice constant has contracted. Whereas (220) peak shifts towards a lower angle for x = 0.06 and is related to an increase in unit cell volume due to interstitial Ti atoms. Further, Rietveld refinement has been employed to estimate the Debye-Waller factor ($B_{iso}$), related to positional disorder. The effect of extra Sb on $B_{iso}$ of the synthesized samples (TiCoSb1+x) has been presented in Fig. 7. A non-monotonic variation of $B_{iso}$ with Sb concentration has been observed and anomaly for x=0.02 has been corroborated with other structural data, viz, FWHM of the XRD peak, unit cell volume, estimated wt% of embedded phases. The Debye equation may be applied to estimate Debye temperature ($\theta_D$) if the harmonic approximation, i.e., cubic and monoatomic structure has been considered. $\theta_D$ and $B_{iso}$ are related with the following equation within harmonic approximation[51]

$$B_{iso} = \left[\frac{6h^2}{MK_B\theta_D}\right]\left[\frac{1}{4} + \left(\frac{T}{\theta_D}\right)^2 \int_0^{\frac{\theta_D}{T}} \frac{x}{e^x - 1} dx\right] \quad (1)$$

where, M and T are mass and temperature respectively. Non-monotonic variation of $\theta_D$ of the prepared samples has been presented in TABLE II and lowest $\theta_D$ ($\theta_D$=357 K) has been observed for TiCoSb1.02 sample. Skovsen et al. have also reported a similar value of $\theta_D$~357.[52] Further, $\theta_D$~417K has been estimated and reported by T. Sekimoto et al. employing measurement of longitudinal and shear velocities using the ultrasonic piles-echo method.[53]

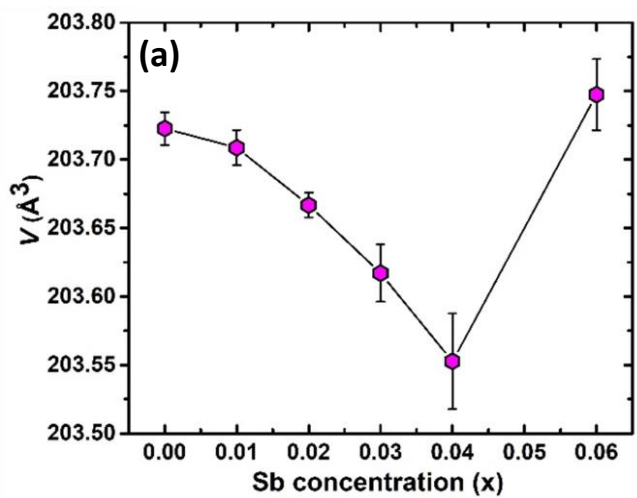 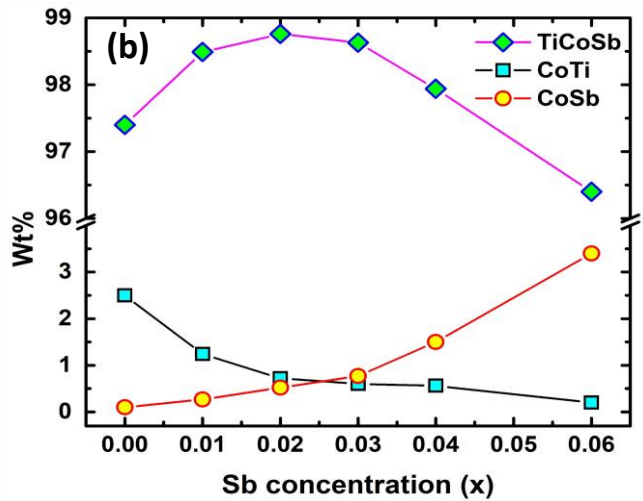

**Fig. 5.** (Color online). Sb concentration dependent (a) unit cell volume and (b) Wt% of TiCoSb and embedded CoTi, CoSb phases, as obtained by employing Rietveld refinement of XRD data of synthesized TiCoSb1+x (x=0.0, 0.01, 0.02, 0.03, 0.04, 0.06) polycrystalline sample.

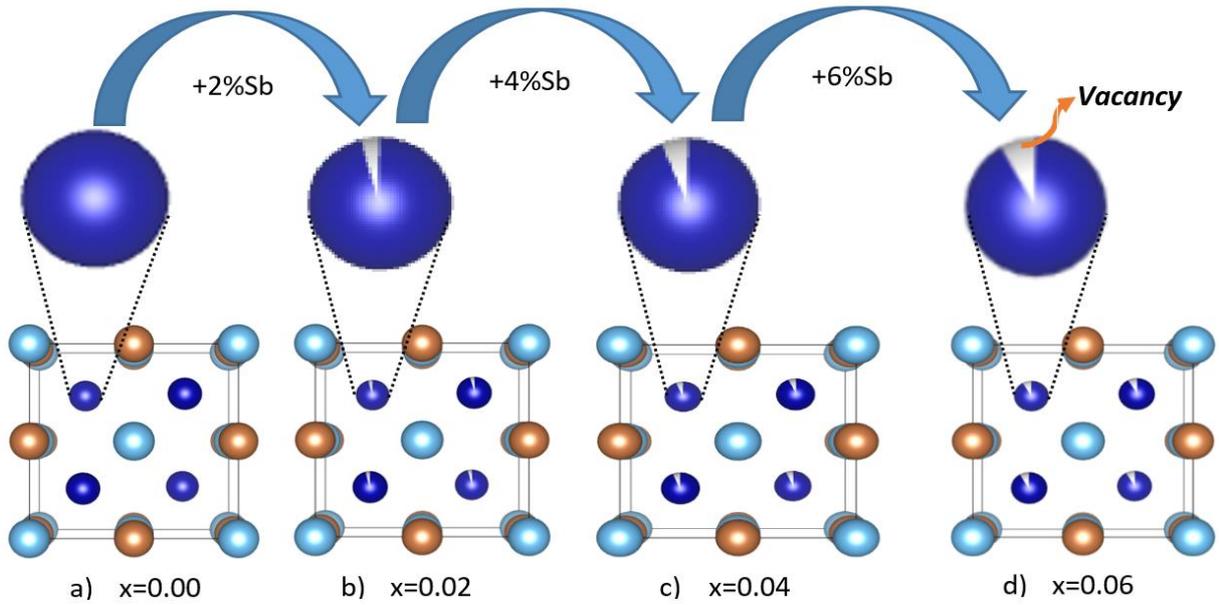

**Fig. 6.** (Color online). Schematic representation of formation of vacancy in TiCoSb1+x (x=0.0, 0.01, 0.02, 0.03, 0.04, 0.06) polycrystalline samples, obtained from Rietveld refinement, using vesta software. Sky blue spheres represent Ti atoms, brown spheres represent Sb atoms and navy blue spheres represent Co atoms. The white region in Co atoms (zooming view) represents vacancy.

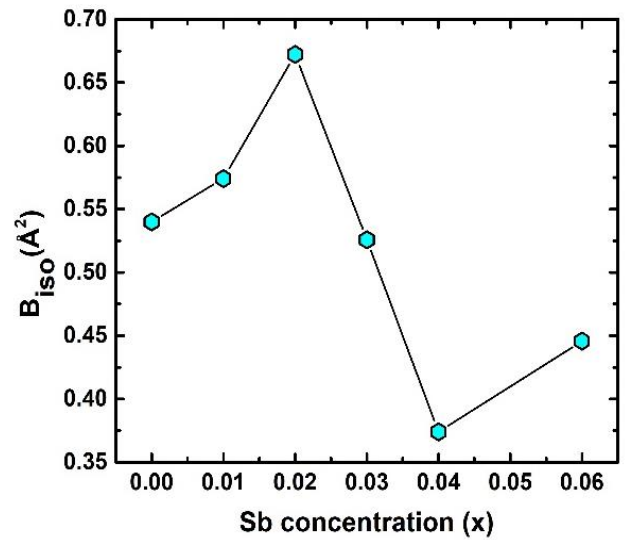

**Fig. 7.** (Color online). Variation of Debye-Waller factor ($B_{iso}$) with Sb concentration of synthesized TiCoSb1+x (x=0.0, 0.01, 0.02, 0.03, 0.04, 0.06) polycrystalline materials. $B_{iso}$ have been obtained from the Rietveld refinement of XRD data.

Non-monotonic variation of ϵ and d of the synthesized samples has been depicted in Fig. 8. The Williamson-Hall method has been employed to estimate ϵ and d using the relation[38]

$$\beta cos\theta = \frac{K\lambda}{d} + 4\varepsilon\, sin\theta. \quad (2)$$

Estimated ϵ for all TiCoSb1+x (x= 0, 0.01, 0.02, 0.03, 0.04, 0.06) samples is negative in nature, indicates compressive strain[54] (Please find supplementary Fig. S2 for Williamson-Hall plot). The variations of ϵ and d with the Sb concentration of the synthesized samples are dissimilar in nature. The magnitude of ϵ is minimum and d is maximum for x=0.02 amid TiCoSb1+x (x= 0, 0.01, 0.02, 0.03, 0.04, 0.06), synthesized samples. It may arises due to the lowest wt% of embedded phases viz, CoTi, CoSb and maximum TiCoSb phase in TiCoSb1.02. However, the highest d for the TiCoSb1.02 sample may be explained in light of the well-known Hall-Petch formula. According to the Hall-Petch formula, the strengthening effect of grain boundaries may be described as [55, 56, 57]

$$\sigma_{GB} = k_y d^{-1/2}. \quad (3)$$

Where $\sigma_{GB}$, d, $k_y$ are yield strength at grain boundaries, crystalline size, a material dependent constant respectively. Lattice mismatch at the grain boundary due to embedded

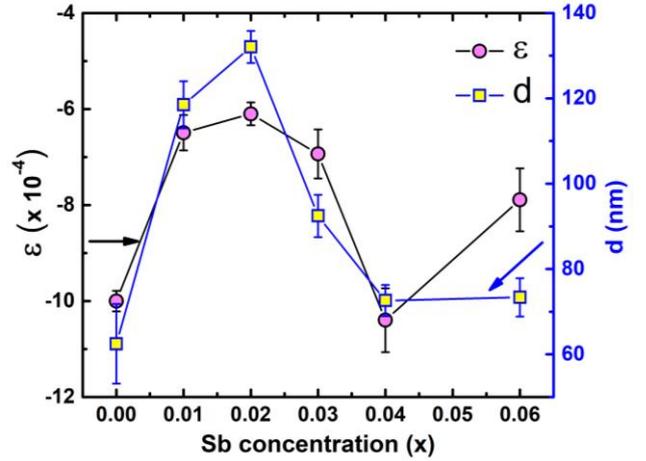

**Fig. 8**. (Color online). Sb concentration dependent lattice strain (ε) and crystalline size (d) of $TiCoSb$1+x (x=0.0, 0.01, 0.02, 0.03, 0.04, 0.06) polycrystalline materials, obtained from x-ray diffraction data using Williamson-Hall equation.

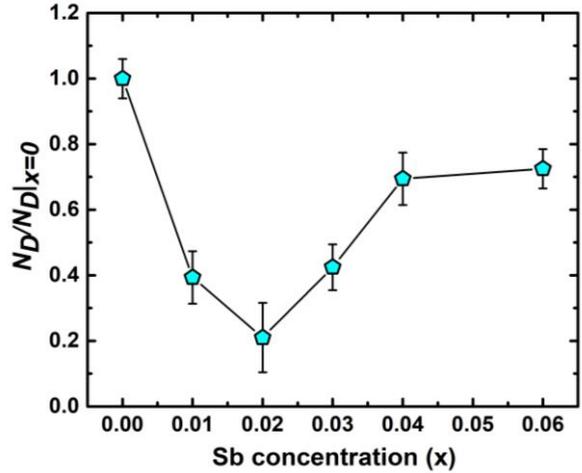

**Fig. 9.** (Color online). Relative dislocation density ($N_D/N_D|x=0$) for the $TiCoSb$1+x (x=0, 0.01, 0.02, 0.03, 0.04, 0.06) polycrystalline samples, estimated from x-ray diffraction data using modified Williamson-Hall equation.

phases may cause low $\sigma_{GB}$ and high d value for TiCoSb1.02 sample.

Further, XRD data may be employed to estimate the dislocation density ($N_D$) in the synthesized materials. The Modified Williamson-Hall plot has been used to carry out the calculation for estimation of $N_D$ in TiCoSb1+x (x= 0, 0.01, 0.02, 0.03, 0.04, 0.06) samples, employing the relation[39, 58]

$$\Delta K = 0.9/\delta + \frac{\pi A^2 B_D^2}{2} N_D^{\frac{1}{2}} K^2 C \pm O(K^4 C^2). \quad (4)$$

K [=$2sin\theta_B/\lambda$] and $\Delta K$ [=$\Delta(2\theta_B)cos\theta_B/\lambda$] parameters have been estimated from XRD data, where $\theta_B$ is the Bragg angle, $\Delta(2\theta_B)$ is the FWHM of the corresponding diffraction peak at $\theta_B$ and λ (=0.15406 nm) is the wavelength of Cu- Kα X-ray.

$N_D$ has been obtained from the slope of '$\Delta K$ versus $K^2C$' curve, where C is average dislocation contrast factor. The '$\Delta K$ versus $K^2C$' plots of the synthesized TiCoSb1+x (x=0.0, 0.01, 0.02, 0.03, 0.04, 0.06) samples have been given in the supplementary information [Fig. S3]. $N_D$ of the synthesized samples (TiCoSb1+x; x=0.0, 0.01, 0.02, 0.03, 0.04, 0.06) has been estimated with respect to the TiCoSb (i.e., x=0.00) sample. Non-monotonic behaviour of relative dislocation density ($N_D/N_D|_{x=0}$) as a function of Sb concentration(x) has been depicted in Fig. 9. $N_D$ increases on either side of the point x = 0.02, lowest $N_D$ has been observed for the TiCoSb1.02 sample. This is in agreement with the previous analysis that the presence of a minimum embedded phase with good crystal quality has been observed for the TiCoSb1.02 sample amid synthesized samples (TiCoSb1+x; x=0.0, 0.01, 0.02, 0.03, 0.04, 0.06). An increase in the wt% of embedded phases in the synthesized samples may enhance the lattice mismatch at the grain boundary, causes an increment in $N_D$.

Temperature dependent S(T) have indicated all the synthesized samples (Fig. 10(a)), TiCoSb1+x (x = 0.0, 0.01, 0.02, 0.03, 0.04, 0.06) are n-type in nature. Negative value of S(T) has indicated majority carriers are electrons. However, pristine TiCoSb should have shown positive S(T) i.e., p-type in nature.[59] Synthesized samples have shown an n-type nature as Fermi surfaces have modified due to embedded phases and defects. The magnitude of S(T) values are lower than most of the previously reported numbers.[23, 60, 61] However, Sekimoto et al. have also reported a lower S(T) value of TiCoSb ~ -150 μV/K at 350 K.[48] An increase in S(T) data beyond 275 K may be attributed to thermal excitation of minority carriers at high temperatures.[62] It has been found that S(T) for TiCoSb (i.e., x=0.00) is nearly zero. And |S(T)| of the synthesized samples,

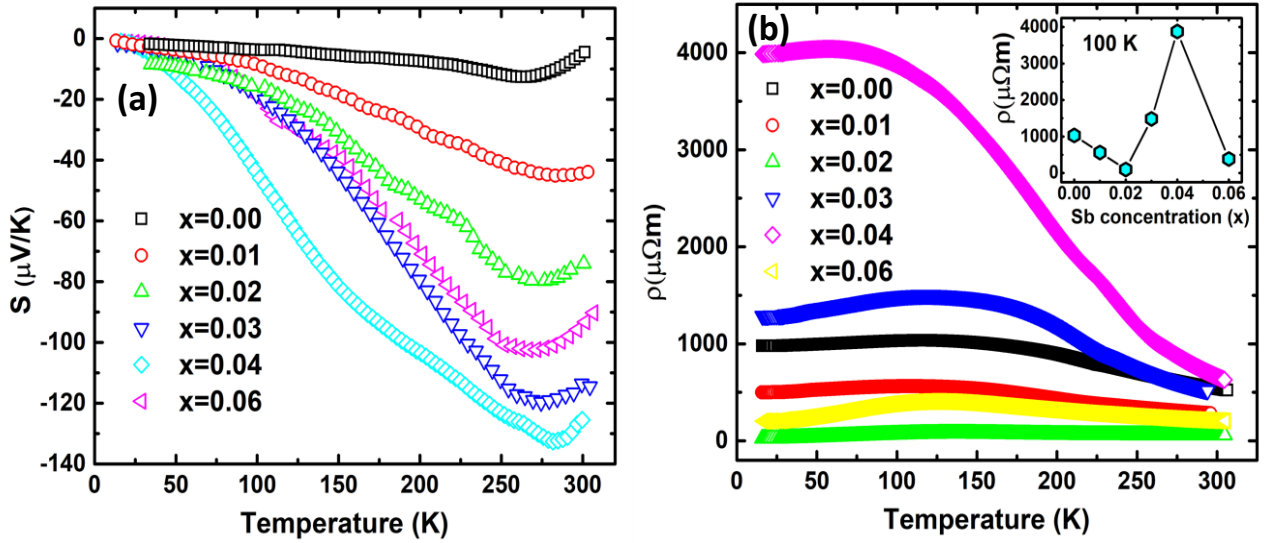

**Fig. 10.** (Color online). Temperature dependent (a) Seebeck coefficient S(T) and (b) electrical resistivity ρ(T) of TiCoSb1+x (x=0.0, 0.01, 0.02, 0.03, 0.04, 0.06) half-Heusler alloy. (Inset of (b) indicates Sb concentration dependent ρ at 100 K).

TiCoSb1+x (0.0<x<0.04) increases with increasing Sb concentration. However, |S(T)| has slightly reduced for TiCoSb1.06.

The non-monotonic behavior of S(T) with Sb concentration may be attributed to the band structure modification owing to the presence of defects (vacancy, interstetial and antisite) and embedded phases (CoTi and CoSb). The thermopower of a semiconductor may be written as[63, 64]

$$|S| = \frac{(\sigma_n|S_n| - \sigma_p|S_p|)}{\sigma_n + \sigma_p}. \quad (5)$$

Here, $S_n/S_p$ ($S_{n,p}$) and $\sigma_n/\sigma_p$ ($\sigma_{n,p}$) are the Seebeck coefficients and the electrical conductivities due to n-type and p-type carriers, respectively. It has been assumed that both types of charge carriers are involved in the conduction process. However, $S_{n,p}$ may be written as, according to the Mott expression[65, 66]

$$|S_{n,p}| = \frac{\pi^2 K_B^2 T}{3e}\left[\frac{1}{n_{n,p}}\frac{dn_{n,p}(E)}{dE} + \frac{1}{\mu_{n,p}}\frac{d\mu_{n,p}(E)}{dE}\right]_{E \approx E_F} \quad (6)$$

here, $n_{n,p}$ is energy dependent n-type or p-type carrier density, $E_F$, e and $K_B$ are Fermi energy, electronic charge and Boltzmann constant. The mobility of n-type or p-type charge carrier has been denoted as $\mu_{n,p}$. Further, energy dependent $n_{n,p}(E)$ is related with DOS, $n_{n,p}(E) = g_{c,v}(E)f(E)$. Here $g_{c,v}$ represents DOS in conduction band/valance band, near Fermi level and f(E) is Fermi distribution function. Thermal variation of S(T) of synthesized TiCoSb1.0 is almost zero, indicates compensated behaviour of charge carriers in the conduction processes, according to the equation 5. However, S(T) for the TiCoSb1+x (0≤x≤0.04) may be explained, according to the Mott expression (equation 6). Theoretical calculation of band structure and DOS suggest that Ti and Co atoms mainly contribute to the conduction and valence bands of TiCoSb, respectively. It should be recalled from the Rietveld refinement data that Co vacancies have incrised with Sb concentration for the synthesized samples (Fig. 6). Hence, Co vacancies in the TiCoSb crystal directly contribute to the DOS in the valance band of TiCoSb. As Co vacancy created, DOS in the valence band near $E_F$ dereases. And energy dependent $n_p(E\sim E_F)$ may decreases, owing to the local decrease in DOS in the valence band ($g_v(E)$) near the $E_F$. It causes decrease in $|S_p|$ in the synthesized samples. Further, carriers near the $E_F$ mostly contribute to the conduction. $\sigma = ne\mu$ indicates that a drop in $n_p(E\sim E_F)$ near $E_F$ may cause a decrease in $\sigma_p$ and a concomitant reduction in $\sigma_{tot}$. Hence, the resultant $|S(T)|$, according to the equation 6 for TiCoSb1+x (0.01≤x≤0.04) have enhanced, provided that $E_F$ of the materials aligns properly. Maximum wt% of CoSb embedded phases along with interstial Sb have been obtained for the TiCoSb1.06 sample. The position of the DOS-peak for CoSb embadded phases exactly matches the position of DOS at the valence band of TiCoSb[67]. The complex interplay of interstitial Sb atom and DOS of CoSb support to increase in DOS at valence band of TiCoSb1.06 sample and results in a slight decrease in S(T).

**TABLE II.** $B_{iso}$, $\theta_D$, $\kappa_e$, $\kappa_{tot}$, $\rho$, S, ZT of TiCoSb1+x (x=0.0, 0.01, 0.02, 0.03, 0.04, 0.06) polycrystalline materials with different extra wt% Sb concentration (x) at room temperature.

| Sb Concentration (x). | $B_{iso}$ (Å$^2$) | $\theta_D$ K | $\kappa_e$ W/mK | $\kappa_{tot}$ W/mK | $\rho_{300K}$ μΩm | $S_{300K}$ μV/K | $ZT_{300K}$ |
|---|---|---|---|---|---|---|---|
| 0.00 | 0.53 | 287 | 0.01339 | 4.945 | 536.54 | -4.5 | 9.6 x10$^{-6}$ |
| 0.01 | 0.57 | 378 | 0.02308 | 4.6185 | 283.42 | -44.5 | 8 x10$^{-4}$ |
| **0.02** | **0.68** | **356** | **0.10422** | **3.94279** | **58.34** | **-76** | **0.025** |
| 0.03 | 0.52 | 391 | 0.01078 | 5.09605 | 517.718 | -116.6 | 0.00723 |
| 0.04 | 0.37 | 447 | 0.00767 | 7.605 | 704.5 | -129.4 | 0.00313 |
| 0.06 | 0.44 | 417 | 0.02725 | 6.1972 | 535.22 | -95.8 | 0.00691 |

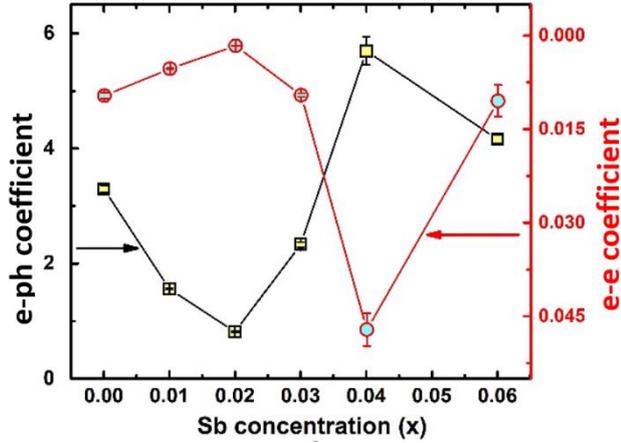

**Fig. 11.** (Color online). Variation of electron-phonon (e-ph) and electron-electron (e-e) scattering coefficient with Sb concentration for synthesized TiCoSb1+x (x=0.0, 0.01, 0.02, 0.03, 0.04, 0.06) polycrystalline samples, obtained by fitting the temperature dependent resistivity data.

$\rho$(T) of all the synthesized samples, TiCoSb1+x (x=0.0, 0.01, 0.02, 0.03, 0.04, 0.06) has been depicted on Fig. 10(a). Non-monotonic thermal variations of $\rho$(T) indicate that TiCoSb1+x (x=0.0, 0.01, 0.02, 0.03, 0.04, 0.06) synthesized samples are semiconducting in nature. $\rho$(T) decreases with increasing Sb concentration for 0.00 ≤ x ≤ 0.02. However, further increase of extra Sb causes increase in $\rho$(T) (0.02 <x≤ 0.04) and highest resistivity is observed for x=0.04 sample. $\rho$(T) at 300 K of the synthesized samples has been presented in TABLE II. $\rho$(T) of TiCoSb1.02 is similar to previously reported value for pristine TiCoSb[48]. Increase in resistivity with Sb concentration for TiCoSb1+x samples may be related with inhrent Co vacancy in the synthesized samples. In the preceding discussion, it has been observed that Co vacancy may leads to decrease in $\sigma_{tot}$ (=$\sigma_n+\sigma_p$) due to change in DOS at valence band. However, embedded phases, defect and disorder also play crucial role in conduction of carrier.[68, 69] Embedded phases in the TiCoSb host phases may act as scattering centre during the conduction processes. In order to get the insight of scattering, e-e and e-ph scattering coefficients have been estimated from the $\rho$(T) data. The $\rho$(T) data has been fitted with equation, $\rho(T)= \rho_0+AT+BT^2$ at low temperature (T<150K), where A and B represent e-ph and e-e scattering coefficients respectively.[57] Non-monotonic variation of e-e and e-ph have been presented in Fig. 11. e-e and e-ph coefficients have decreased for 0.00<x<0.02 in TiCoSb1+x samples and a sudden drop has been observed for TiCoSb1.02 sample. It important to mention that maximum wt% of TiCoSb phase is obtained for TiCoSb1.02 and crystal quality increases for the TiCoSb1.02 sample. The decrease in wt% of TiCoSb phase and increase in Co vacancy along with embedded phases increase the scatttering center of the synthesized samples, TiCoSb1+x (0.02<x≤0.0 and 0.02<x≤0.04). Highest wt% of TiCoSb phase is obtained for TiCoSb1.02 sample and correspondingly lowest ρ(T) is observed. Embedded phases and Co vacancy increase for x=0.03 and 0.04, corroborated with increase in ρ(T). Sudden drop in ρ(T) for TiCoSb1.06 may be related with change in the band structure and formation of impurity band, owing to excessive impurity phases and interstitial atoms.

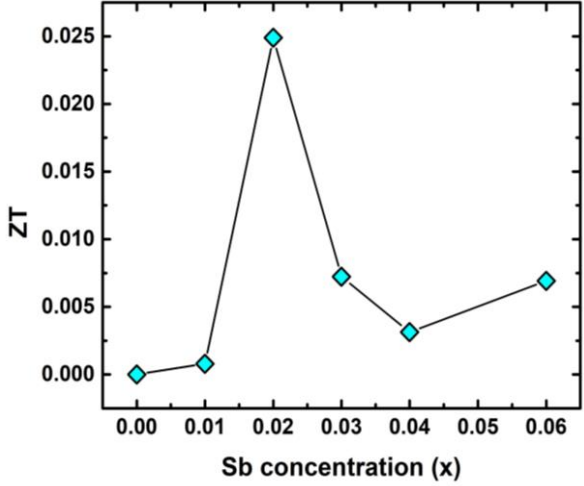

**Fig. 12.** (Color online). Room temperature ZT of TiCoSb1+x (x=0.0, 0.01, 0.02, 0.03, 0.04, 0.06) half-Heusler alloy.

Power factor, PF ($S^2\sigma$) of the synthesized samples have been estimated using S-T and ρ-T data. 'PF versus temperature' data of TiCoSb1+x are depicted in Fig. S4 [Supplementary information]. Non-monotonic thermal variation of power factor is observed. Highest power factor is obtained for TiCoSb1.02 at 300K.

In order to get the flavour of change in ZT due to embedded phases and defects, thermal conductivity κ (κ=$\kappa_e$ + $\kappa_L$, $\kappa_e$ =electronic thermal conductivity and $\kappa_L$= lattice thermal conductivity) has been estimated. $\kappa_e$ is estimated using well known Wiedemann Franz law, $\kappa_e = L\sigma T$ (L is the Lorenz number). Further, L has been calculated from the temperature dependent S(T) data, employing the relation $L = [1.5 + exp(-|S|/116)] \times 10^{-8}$.[70] $\kappa_L$ is directly related with the phonon scattering mechanism viz, phonon-phonon scattering due to defects. Heat conduction is performed by acoustic phonons and Umklapp process is dominant in the phonon scattering at high temperature. However, according to Slack, $\kappa_L$ may be written in the limit Umklapp scattering processes[71]

$$\kappa_L = \Lambda \frac{\bar{M}\theta_D^3\delta}{\gamma_G^2 n^{2/3} T} \qquad (7)$$

where, $\bar{M}$, $\delta^3$ and n are average atomic mass in the unit cell, volume per atom and number of atoms in the primitive cell respectively. There are 12 atoms per unit cell of TiCoSb i.e, n=12 and $\Lambda \sim 3.1 \times 10^{-6}$ (a physical constant). Value of Gruneisen parameter $\gamma_G$ of pristine TiCoSb has been considered to estimate $\kappa_L$ of the synthesized samples using equation 7. It's important to note that $\gamma_G$ is almost temperature independent and depends on stoichiometry of the constituent elements. In this study, we are dealing with TiCoSb material with very small variation of Sb in the stoichiometry. Hence, $\gamma_G$ of TiCoSb i.e, 2.13 has been considered for calculation of $\kappa_L$ using equation 7.[72] ZT of the synthesized samples are estimated using equation, ZT=$S^2\sigma/\kappa$. Value of $\theta_D$, estimated using Rietveld refinement and $\kappa_L$ of the synthesized samples have been presented in TABLE II. Fig. 12 represents dependency of ZT on extra Sb concentration in TiCoSb1+x at room temperature. It crucial to mention that a drastic change in ZT has been observed for TiCoSb1.02 sample. Maximum ZT~0.025 has been observed for TiCoSb1.02 sample, which is nearly 4 to 5 times higher than the previously reported values for the prestine TiCoSb at room temperature.[20, 22, 34, 35, 36]

**V. Conclusion**

In conclusion, effect of embedded phases and defects on structural, resistive and thermopower of TiCoSb HH polycrystalline material have been reported. Theoretical calculations have been employed to estimate the DOS, band structure and formation energy for vacancy, anti-site and interstial defects. TiCoSb1+x (x=0.0, 0.01, 0.02, 0.03, 0.04, 0.06), six samples have been synthesized using solid state reaction method, followed by arc-melting. Excess Sb has been added to TiCoSb stoichiometry to compensate the evaporated Sb, create vacancy and embedded phases, during arc-melting of the constituient elements. Mix-phase Rietveld refinement anslysis of the XRD data clearly reveal the presence of CoTi and CoSb phases. Optimization of CoTi and CoSb phaases and maximum wt% of TiCoSb are observed for TiCoSb1.02. A distinct slope change is observed at 2% extra Sb concentration for 'wt% of TiCoSb' versus 'Sb concentration' graph. Co vacancy increases with Sb concentration, obtained from Rietveld refinement and schematecally presented using VESTA software. $\theta_D$ employing $B_{iso}$, obtained from XRD data of the synthesized samples has been estimated and $\theta_D$ of TiCoSb1.02 matches with reported value. Lattice strain (ε) and grain size (d) have been estimated by Williamson-Hall plots. Minimum ε and maximum d are obtained for 2% extra Sb contained sample, due to improved crystal quality and maximum TiCoSb phase. It also supports the minimum $N_D$ for TiCoSb1.02 sample. S(T) and ρ(T) are related with modification of DOS owing to the Co vacancy and embedded phases, explained by Mott equation. e-e and e-ph scattering coefficients are correlated with the defects and the embedded phases. PF has been estimated using ρ(T) and S(T) data. In order to get the flavour of ZT, $κ_L$ and $κ_e$ are calculated from XRD and resistivity data, using temperature dependent Lorentz number, obtained from S(T). However, final conclusion may be drawn after the detail analysis of the experimental data of κ. ZT of TiCoSb strongly depends on extra Sb added to stoichiometry during synthesis. It has been observed that ZT of TiCoSb1.02 sample is 4 to 5 times greater than previously reported value for TiCoSb at room temperature. Corroborated structural and transport data reveal that the best pristine phase of TiCoSb and greater TE properties of TiCoSb is obtained for 2% extra Sb contained sample. Hence, 2% extra Sb in TiCoSb stoichiometry is sufficient to compensate the loss of Sb during synthesis. And resistivity and thermopower of TiCoSb are influenced due to the modification in DOS near the Fermi surface by defects and embedded phases.

## Acknowledgments

This work is supported by the Science and Engineering Research Board (SERB), Government India in the form of sanctioning research project (File Number: EEQ/2018/001224). Author SM is thankful to CSIR, India for providing Research Fellowships.

*Email: kartick.phy09@gmail.com